\documentclass[submission, Phys]{SciPost}
\usepackage{graphicx}
\usepackage{dcolumn}
\usepackage{bm}
\usepackage{upgreek}
\usepackage{hyperref}
\usepackage{subcaption}
\usepackage{multirow}

\usepackage[utf8]{inputenc} 
\usepackage[T1]{fontenc} 	
\usepackage[english]{babel} 

\usepackage[bitstream-charter]{mathdesign}
\usepackage{geometry} 		
\usepackage{amsmath} 		
\usepackage{mathtools} 		
\usepackage{float} 			
\usepackage{graphicx} 		
\usepackage{tabularx} 		
\usepackage{booktabs} 		
\usepackage{color, xcolor} 	
\usepackage{pdfpages} 		
\usepackage{extarrows} 		
\usepackage{multirow} 		
\usepackage{multicol} 		
\usepackage{adjustbox}
\usepackage{enumitem} 		
\usepackage{xspace} 		
\usepackage{stackrel} 		
\usepackage{tikz} 			
\usepackage{braket} 		
\usepackage{bm} 			
\usepackage{tensor} 		
\usepackage{slashed} 		
\usepackage{siunitx} 		
\usepackage{lastpage} 		
\usepackage{cite} 			
\usepackage[normalem]{ulem} 
\usepackage{fontawesome} 	
\usepackage{tocloft} 		
\usepackage{titlesec} 		
\usepackage{doi} 			
\usepackage{hyperref} 		
\usepackage[most]{tcolorbox} 					
\usepackage[nameinlink, capitalize]{cleveref} 	
\usepackage[nottoc, notlot, notlof]{tocbibind} 	
\usepackage[ruled, vlined]{algorithm2e} 		
\usepackage{makecell}
\usepackage[makeroom]{cancel}

\DeclareSymbolFont{usualmathcal}{OMS}{cmsy}{m}{n}
\DeclareSymbolFontAlphabet{\mathcal}{usualmathcal}

\makeatletter
\def\BState{\State\hskip-\ALG@thistlm}
\makeatother

\makeatletter
\@ifundefined{pdfoutput}{}{\DeclareGraphicsRule{*}{mps}{*}{}}
\makeatother

\makeatletter
\DeclareRobustCommand*{\bfseries}{%
   \not@math@alphabet\bfseries\mathbf
   \fontseries\bfdefault\selectfont
   \boldmath
}
\makeatother

\definecolor{EmeraldGreen}{HTML}{1ea78d}
\definecolor{EnglishRed}{HTML}{b02427}
\hypersetup{
	pdftitle={BitHEP},
	pdfauthor={Krause et al.},
	colorlinks=true, 			
	linkcolor=EnglishRed, 	
	citecolor=EmeraldGreen, 	
	urlcolor=EmeraldGreen 	
} 

\SetArgSty{textnormal}
\SetKwComment{Comment}{{\small\#}~}{}
\SetCommentSty{mycommfont}

\setitemize{itemsep=2pt,topsep=2pt,parsep=0pt,partopsep=0pt,leftmargin=*}
\setenumerate{itemsep=0pt,topsep=2pt,parsep=0pt,partopsep=0pt,labelindent=3pt,leftmargin=*}
\usepackage{amsmath}
 
\usepackage{amsthm} 		
\theoremstyle{definition}
 
\definecolor{red_cb}{HTML}{e41a1c}
\definecolor{blue_cb}{HTML}{377eb8}
\definecolor{green_cb}{HTML}{4daf4a}
\definecolor{purple_cb}{HTML}{984ea3}
\definecolor{orange_cb}{HTML}{ff7f00}

\newcommand\geant{\textsc{Geant}4\xspace}

\newcommand{\eg}{\text{e.g.}\;}
\newcommand{\ie}{\text{i.e.}\;}

\newcommand{\mwith}{\text{with}}

\newcommand{\sign}{\operatorname{sign}} 	
\newcommand{\round}{\operatorname{round}}	

\newcommand{\Attention}{\operatorname{Attention}}

\newcommand{\fastjet}{\textsc{FastJet}\xspace}

\newcommand{\caloinn}{\textsc{CaloINN}\xspace}
\newcommand{\calodream}{\textsc{CaloDREAM}\xspace}
\newcommand{\bitnet}{\textsc{BitNet}\xspace}

\newcommand{\arXiv}[2][]{%
	\ifthenelse{\equal{#1}{}}%
	{\href{http://arxiv.org/abs/#2}{arXiv:#2}}%
	{\href{http://arxiv.org/abs/#2}{arXiv:#2~[#1]}}}

\def\slashchar#1{\setbox0=\hbox{$#1$}           
   \dimen0=\wd0                                 
   \setbox1=\hbox{/} \dimen1=\wd1               
   \ifdim\dimen0>\dimen1                        
      \rlap{\hbox to \dimen0{\hfil/\hfil}}      
      #1                                        
   \else                                        
      \rlap{\hbox to \dimen1{\hfil$#1$\hfil}}   
      /                                         
   \fi}

\newcommand{\tikznode}[2]{%
\ifmmode%
\tikz[remember picture,baseline=(#1.base),inner sep=0pt] \node (#1) {$#2$};%
\else
\tikz[remember picture,baseline=(#1.base),inner sep=0pt] \node (#1) {#2};%
\fi}

\def\mathswitchr#1{\relax\ifmmode{\mathrm{#1}}\else$\mathrm{#1}$\xspace\fi}
\def\mathswitch#1{\relax\ifmmode#1\else$#1$\xspace\fi}

\newcommand{\Pq}{\mathswitch q}
\newcommand{\Pqbar}{\mathswitch{\bar q}}
\newcommand{\PW}{\mathswitchr W}

\newcommand{\PZ}{\mathswitchr Z}

\newcommand{\Pg}{\mathswitchr g}

\newcommand{\Pnu}{\nu}
\newcommand{\Pnubar}{\bar \nu}

\newcommand{\Pp}{\mathswitchr p}

\graphicspath{{./figs/}}

\begin{document}
\begin{flushright}
    \vspace*{-1cm}
    HEPHY-ML-25-02
    \vspace*{0.5cm}
\end{flushright}

\begin{center}{\Large \textbf{
BitHEP --- The Limits of Low-Precision ML in HEP
}}\end{center}

\begin{center}\textbf{
Claudius Krause\textsuperscript{1},
Daohan Wang\textsuperscript{1} and
Ramon Winterhalder\textsuperscript{2}
}\end{center}

\begin{center}
{\bf 1} HEPHY, Austrian Academy of Sciences (OeAW), Vienna, Austria
\\
{\bf 2} TIFLab, Universit\`a degli Studi di Milano \& INFN Sezione di Milano, Italy

\end{center}

\begin{center}
\today
\end{center}

\section*{Abstract}
{\bf The increasing complexity of modern neural network architectures demands fast and memory-efficient implementations to mitigate computational bottlenecks. In this work, we evaluate the recently proposed \bitnet architecture in HEP applications, assessing its performance in classification, regression, and generative modeling tasks. Specifically, we investigate its suitability for quark-gluon discrimination, SMEFT parameter estimation, and detector simulation, comparing its efficiency and accuracy to state-of-the-art methods. Our results show that while \bitnet consistently performs competitively in classification tasks, its performance in regression and generation varies with the size and type of the network, highlighting key limitations and potential areas for improvement.}

 
\vspace{10pt}
\noindent\rule{\textwidth}{1pt}
\tableofcontents
\noindent\rule{\textwidth}{1pt}
\vspace{10pt}

\clearpage

\section{Introduction}
\label{sec:intro}

The upcoming high-luminosity phase of the LHC (HL-LHC) will push the boundaries of precision measurements and new physics searches, necessitating unprecedented advances in computational methods. The ability to efficiently generate and analyze vast amounts of collision data is essential for maximizing the scientific potential of the HL-LHC. To this end, modern machine learning (ML) techniques have become indispensable tools in high-energy physics, enabling more precise and efficient modeling of complex physical processes~\cite{Butter:2022rso, Plehn:2022ftl}. 

Machine Learning has already demonstrated its utility in various HEP domains, accelerating various aspects of a sophisticated simulation and analysis chain. In particular, for event generation, deep learning models are widely used for tasks such as scattering amplitude evaluations~\cite{Bishara:2019iwh,Badger:2020uow,Aylett-Bullock:2021hmo,Maitre:2021uaa,Winterhalder:2021ngy,Badger:2022hwf,Maitre:2023dqz,Bahl:2024gyt}, phase-space sampling~\cite{Bendavid:2017zhk,Klimek:2018mza,Chen:2020nfb,Gao:2020vdv,Bothmann:2020ywa,Gao:2020zvv,Danziger:2021eeg,Heimel:2022wyj,Janssen:2023ahv,Bothmann:2023siu,Deutschmann:2024lml,Heimel:2024wph}, parton shower generation~\cite{deOliveira:2017pjk,Andreassen:2018apy,Bothmann:2018trh,Dohi:2020eda,Buhmann:2023pmh,Leigh:2023toe,Mikuni:2023dvk,Buhmann:2023zgc}, and detector
simulations~\cite{Paganini:2017hrr,deOliveira:2017rwa,Paganini:2017dwg,Erdmann:2018kuh,Erdmann:2018jxd,Belayneh:2019vyx,Buhmann:2020pmy,Buhmann:2021lxj,Krause:2021ilc,
  ATLAS:2021pzo,Krause:2021wez,Buhmann:2021caf,Chen:2021gdz,
  Mikuni:2022xry,ATLAS:2022jhk,Krause:2022jna,Cresswell:2022tof,Diefenbacher:2023vsw,
Hashemi:2023ruu,Xu:2023xdc,Diefenbacher:2023prl,Buhmann:2023bwk,Buckley:2023daw,Diefenbacher:2023flw,Ernst:2023qvn,Favaro:2024rle,Buss:2024orz,Quetant:2024ftg
}. Similarly, for inference and analysis, ML-based approaches have been extensively used for numerous applications, including for instance quark-gluon tagging~\cite{ATL-PHYS-PUB-2017-017,Komiske:2016rsd,Cheng:2017rdo,Stoye:DLPS2017,Chien:2018dfn,Moreno:2019bmu,Kasieczka:2018lwf,1806025,Lee:2019ssx,Lee:2019cad,Dreyer:2020brq,Romero:2021qlf,Filipek:2021qbe,Dreyer:2021hhr,Bright-Thonney:2022xkx,CrispimRomao:2023ssj,Athanasakos:2023fhq,He:2023cfc,Shen:2023ofd,Dolan:2023abg,Blekman:2024wyf,Sandoval:2024ldp,Wu:2024thh,Tagami:2024gtc,Brehmer:2024yqw,Geuskens:2024tfo}, and parameter estimation~\cite{Andreassen:2019nnm,Stoye:2018ovl,Hollingsworth:2020kjg,Brehmer:2018kdj,Brehmer:2018eca,Brehmer:2019xox,Brehmer:2018hga,Cranmer:2015bka,Andreassen:2020gtw,Coogan:2020yux,Flesher:2020kuy,Bieringer:2020tnw,Nachman:2021yvi,Chatterjee:2021nms,NEURIPS2020_a878dbeb,Mishra-Sharma:2021oxe,Barman:2021yfh,Bahl:2021dnc,Arganda:2022qzy,Kong:2022rnd,Arganda:2022zbs,Butter:2022vkj,Neubauer:2022gbu,Rizvi:2023mws,Heinrich:2023bmt,Breitenmoser:2023tmi,Erdogan:2023uws,Morandini:2023pwj,Barrue:2023ysk,Espejo:2023wzf,Heimel:2023mvw,Chai:2024zyl,Chatterjee:2024pbp,Alvarez:2024owq,Diaz:2024yfu,Mastandrea:2024irf,JETSCAPE:2024cqe,Bahl:2024meb,Maitre:2024hzp,Heimel:2024drk}.

However, many of these approaches face significant scalability challenges when deployed in realistic experimental environments. For example, real-time triggering, event reconstruction, and particle tracking applications require ultra-fast inference on resource-constrained hardware such as field-programmable gate arrays (FPGAs)~\cite{Duarte:2018ite,Summers:2020xiy,Iiyama:2020wap, Heintz:2020soy,Aarrestad:2021zos,Hong:2021snb,Migliorini:2021fuj,Govorkova:2021utb,Elabd:2021lgo,Sun:2022bxx,Khoda:2022dwz,Carlson:2022dgb,Abidi:2022ogh,Herbst:2023lug,Coccaro:2023nol,Neu:2023sfh,Borella:2024mgs,Serhiayenka:2024han}, where the complexity of deep neural networks poses a significant bottleneck. Other tasks, like detector simulation or recently-proposed foundation models~\cite{Tani:2025osu,Mikuni:2025tar,Amram:2024fjg,Ho:2024qyf,Wildridge:2024yeg,Leigh:2024ked,Mikuni:2024qsr,Harris:2024sra,Birk:2024knn} require larger and larger networks, which in turn need more disk space to be stored and energy to run. 

A promising avenue for addressing these scalability challenges is model quantization~\cite{DBLP:journals/corr/HanPTD15,han2016deepcompressioncompressingdeep,DBLP:journals/corr/LinTA15,DBLP:journals/corr/CourbariauxBD15,DBLP:journals/corr/CourbariauxB16,DBLP:journals/corr/LiL16,DBLP:journals/corr/abs-1710-09282,DBLP:journals/corr/abs-1711-00215,DBLP:journals/corr/abs-1911-13299,Jin:2023xts}, where neural networks are compressed to lower-bit representations while retaining competitive accuracy. In natural language processing (NLP) and large language models (LLMs), recent research has explored weight matrices with only a few discrete states, significantly reducing memory and computational requirements~\cite{2023arXiv231011453W, 2024arXiv240217764M,10146255}. Similar techniques were investigated in HEP a few years ago, mainly for classification tasks~\cite{DiGuglielmo:2020eqx,Hawks:2021ruw}. However, the potential of such approaches for more general HEP tasks, such as generative modeling~\cite{Rehm:2021zow}, remains largely unexplored. Given these astonishing performances at reduced resource consumption in these initial studies, we expect more development towards broadly-available hardware, dedicated for low-precision computations in the near future, further motivating our study~\cite{Suarez:2025cnd}. 

In this work, we evaluate the recently proposed \bitnet architecture~\cite{2023arXiv231011453W, 2024arXiv240217764M} for key HEP applications, focusing on three fundamental tasks: (i) quark-gluon tagging with a Particle Dual Attention Transformer (P-DAT)\cite{He:2023cfc}, (ii) estimating EFT parameters through regression with SMEFTNet~\cite{Chatterjee:2024pbp}, and (iii) generative modeling for detector simulation using \caloinn~\cite{Ernst:2023qvn} and \calodream~\cite{Favaro:2024rle}. Our goal is to assess whether quantization-aware training (QAT) of \bitnet-based models can achieve accuracy comparable to conventional high-precision networks while reducing computational overhead.

This paper is structured as follows: In Sec.~\ref{sec:bitnet_basics}, we introduce the \bitnet architecture and outline our experimental setup. Afterwards, in Secs.~\ref{sec:classification} -- \ref{sec:gen_exp}, we present our results across the three application domains. Finally, we discuss the implications of our results for future ML-based solutions in HEP and conclude in Sec.~\ref{sec:outlook}.

\clearpage
\section{The \bitnet architecture}
\label{sec:bitnet_basics}

We employ \bitnet, originally designed for LLMs~\cite{2023arXiv231011453W,2024arXiv240217764M}, to various HEP applications. In particular, our approach leverages BitLinear layers to efficiently manage memory usage and computational demands. \bitnet employs either binary or ternary weights with low precision and quantized inputs during forward pass while maintaining high precision for optimizer states and gradients throughout training. This balance ensures both scalability and stability, making it well-suited for the intensive computational requirements of HEP. After training, the network weights can be saved at low precision and only a single floating-point value per layer ($\beta$, to be introduced below) needs to be stored in addition. In contrast to post-training quantization, this quantization-aware training (QAT) approach has the model in low precision already during the training process, which typically leads to better accuracy.

\subsection{BitLinear layer}

\begin{figure}[bht!]
    \centering
    \includegraphics[width=0.45\textwidth]{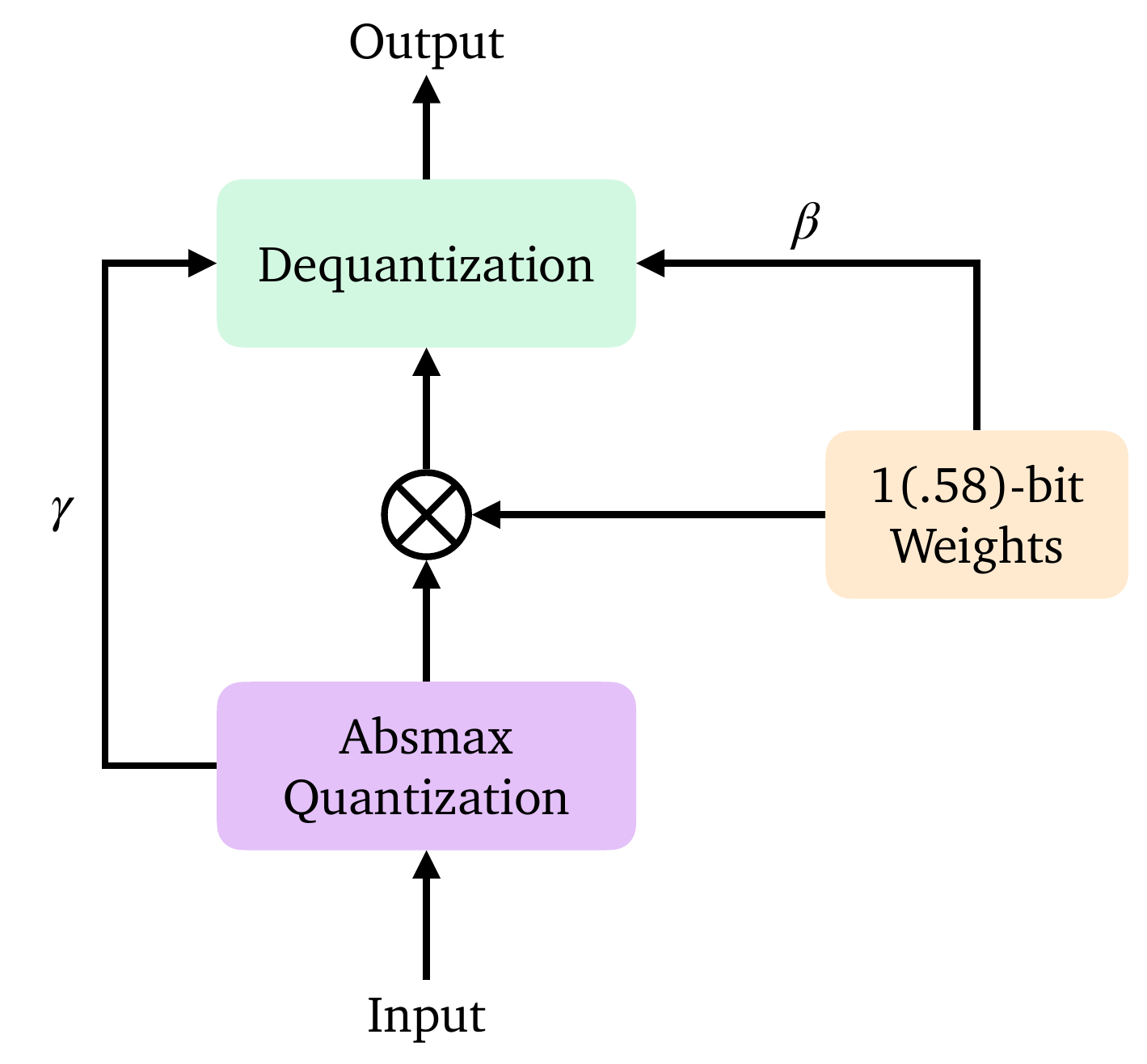}
    \caption{Illustration of the BitLinear layer.}
    \label{fig:bitlinear}
\end{figure}

As illustrated in Fig.~\ref{fig:bitlinear}, both layers begin by mapping the trainable weights $\theta$ onto a quantized representation $\theta_q$. The choice of quantization depends on the acceptable trade-off between precision and efficiency. The standard approach~\cite{2023arXiv231011453W} uses binary weights, $\theta_q \in \{+1,-1\}$, while an extended version~\cite{2024arXiv240217764M} allows for ternary weights, $\theta_q \in \{+1, 0, -1\}$, improving feature filtering. Both weight quantizations are parametrized as
\begin{align}
    \text{binary (1-bit):}& & 
    \theta_q &= \sign\left(\theta - \langle \theta \rangle\right), \notag\\
    \text{ternary (1.58-bit):}& &
    \theta_q &= \max\left(-1, \min\left(1, \round\left(\frac{\theta}{\beta}\right)\right)\right),\qquad \mwith \quad \beta = \left\langle |\theta| \right\rangle.
    \label{eq:quant_weights}
\end{align}
Here, $\left\langle \cdot \right\rangle$ indicates the mean. Beyond weight quantization, both layers apply absmax quantization to the input, ensuring $b$-bit precision. This method scales the input within $[-Q_b, Q_b]$ ($Q_b=2^{b-1}$) by normalizing them against the absolute maximum input value:
\begin{align}
    x_q = \max\left(-Q_b, \min\left(Q_b, \round\left(\frac{x Q_b}{\gamma}\right)\right)\right),
    \qquad \mwith \quad \gamma = \max(|x|)\;.
    \label{eq:quant_activation}
\end{align}
Afterwards, we perform the matrix multiplication between the quantized weights and the input. The output is then rescaled and dequantized using $\{\beta, \gamma\}$ to restore its original precision, allowing the BitLinear layer to be parametrized as
\begin{align}
    y = \theta_q x_q \times \frac{\beta \gamma}{Q_b}.
    \label{eq:quant_output}
\end{align}
The key part of Eq.~\eqref{eq:quant_output} is the matrix multiplication of $\theta_q x_q$. Instead of expensive floating point multiplications, followed by a sum, the quantization of the weights just results in a sign for $x_q$, so $\theta_q x_q$ collapses to a sum without floating point multiplication. 

In all our experiments, we only employ ternary (1.58-bit) quantized weights and use an 8-bit input quantization, \ie $b=8$ and $Q_b=128$. Hence, for the sake of readability, we will always refer to BitLinear or \bitnet in the following without explicitly mentioning 1.58-bit anymore.

\subsection{Implementation and code availability}

The code and data for this paper can be found on GitHub\footnote{see \href{https://github.com/ramonpeter/hep-bitnet}{https://github.com/ramonpeter/hep-bitnet}}, providing readers with the resources needed to reproduce our results or adapt the BitLinear layer for their own implementations.

Currently, our implementation of BitLinear follows a pseudo-quantized approach. While weights are constrained to binary or ternary values, matrix multiplications are performed in full precision (32 or 64 bits). This is because current GPU hardware does not yet support efficient computations at such low precisions, but only to 4-bit integers in some cases~\cite{nvidia_precision}. However, specific computation kernels for CPUs are currently in development~\cite{microsoft_bitnet,2025arXiv250211880W}. A proper study on timing and resource consumption of \bitnet is therefore not part of the present work and remains an avenue for future research. Our pseudo-quantized approach is, however, sufficient for this work, as we focus on evaluating performance metrics rather than computational efficiency.

\subsection{Computational resource requirements}

The relative timing of different numerical operations, \eg floating-point operations (FLOPs) or integer operations (IntOPs), is highly hardware dependent. Since we want to factorize out this component from our analysis, we report on the number of required operations instead. This will serve as a proxy for the resource complexity. 
The number of operations of a regular linear layer, including a bias term but excluding an activation function, is
\begin{align}
N_\text{OPs}=N_\text{FLOPs}= b \times n_\text{out} \times 2 n_\text{in}\;.
\label{eq:FLOPS.linear}
\end{align}
where $b$ is the batch size and $n_\text{in/out}$ denotes the size of the input/output.
In the BitLinear and BitLinear158 layer, the number of operations is 
\begin{alignat}{3}
N_\text{OPs}&\le N_\text{FLOPs} + N_\text{IntOPs} + N_\text{SignOPs} \qquad \mwith &\quad \quad
N_\text{FLOPs} &= b\times \left(3n_\text{out} +n_\text{in} + 1\right)\notag \\
&& N_\text{IntOPs} &= b\times n_\text{out}\times (n_\text{in} -1) \notag\\
&& N_\text{SignOPs} &= b\times n_\text{out}\times n_\text{in}\;.
\label{eq:FLOPS.bitlinear}
\end{alignat}
The ``less than'' factor arises because zero weights in the ternary quantization do not contribute to the sum, so the number of IntOPs is smaller than the sum over all elements. The three FLOPs come from the scaling with $\beta$, $\gamma$, and the bias term. The term independent of the output size comes from the initial quantization of the input vector, which is performed once for all outputs. 

The timing comparisons among FLOP, SignOP, and IntOP are hardware dependent, with dedicated hardware obviously much faster than the general-purpose hardware usually available. In general, the SignOP is basically free, as it is a copy/bitflip/skip operation for weights $\{1, -1, 0\}$ on any hardware. The IntOP depends on the kind of FLOP one compares against and the kind of integer used. Here, we have integer additions, which are even cheaper than integer multiplications. An estimated difference between this IntOP and a FLOP is of the order of a factor 10-30. 

\begin{figure}[tb!]
    \centering
    \includegraphics[width=0.6\linewidth]{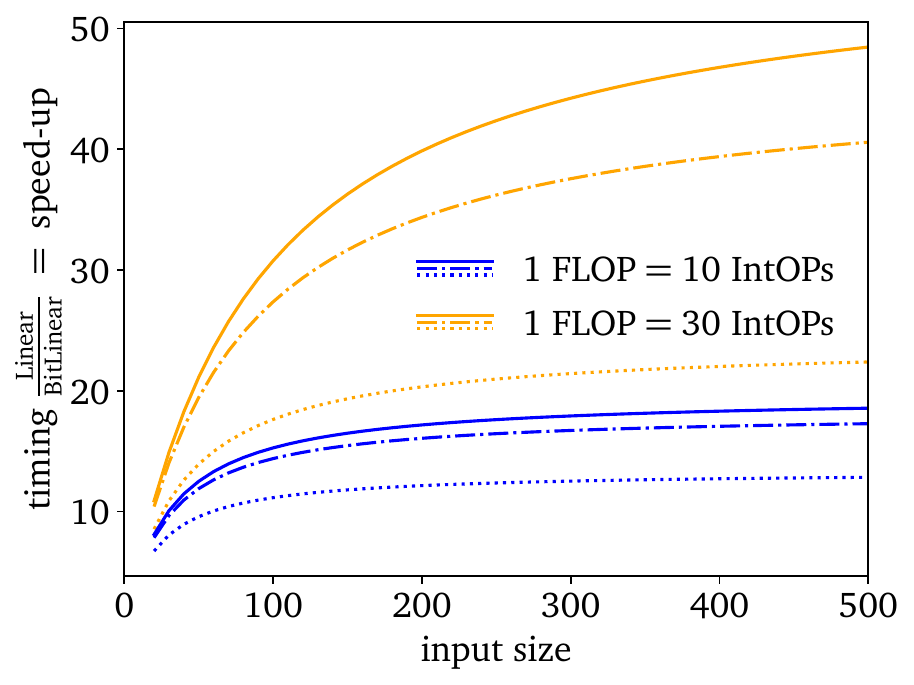}
    \caption{Relative runtime of a regular linear layer and the BitLinear layer as a function of the layer input size for two different assumptions on the relative speed of FLOPs and IntOPs and for three different output sizes: 20 (dotted), 100 (dash-dotted), and 500 (solid).}
    \label{fig:speed-up}
\end{figure}

Figure~\ref{fig:speed-up} shows the ratio of the two scalings for the conservative assumption 1 FLOP = 10 IntOPs and 1 FLOP = 30 IntOPs, for different input and output sizes. In all cases, we assume vanishing costs for SignOPs and the maximal number of IntOPs (i.e. no zero weights that reduce the size of the sum).  

Based on these considerations, it is clear that two networks with the same number of trainable parameters and the same fraction of quantization can still achieve very different speed-ups depending on the number of layers and their widths within the network.  

\clearpage
\section{Classification: Quark-gluon tagging}
\label{sec:classification}

In this section, we employ \bitnet for a quark-gluon discrimination task, utilizing the Particle Dual Attention Transformer (P-DAT)~\cite{He:2023cfc}. This architecture is specifically designed to capture both local particle-level information and global jet-level correlations.\medskip 

To evaluate the performance of the default P-DAT and its quantized counterpart, P-DAT-Bit, we utilize the Quark-Gluon benchmark dataset~\cite{Komiske:2018cqr}, which consists of:
\begin{align}
\begin{split}
   \text{Signal}:& \qquad \Pq \Pqbar \to \PZ(\to \Pnu \Pnubar) + \Pq_{uds},\\
   \text{Background}:& \qquad \Pq \Pqbar \to \PZ(\to \Pnu \Pnubar) + \Pg\;.
\end{split}
\end{align}
Jet clustering is performed using the anti-$k_T$ algorithm with a radius parameter of $R = 0.4$ using \fastjet~\cite{Cacciari:2011ma}. We select only jets with transverse momentum $p_\text{T} \in$ [500, 550] GeV and rapidity $|y| < 1.7$ for further analysis. The dataset contains not only the four-momenta of each particle but also particle identification labels, including electron, muon, photon, as well as individual categories for the different charged and neutral hadrons, \ie the full PID parameterization is used.

The dataset is divided into 1.6M training events, 200k validation events, and 200k test events. Our study focuses on the leading 100 constituents per jet, leveraging their four-momentum components and particle identification labels as input features for training. For jets containing fewer than 100 particles, we apply zero-padding to maintain uniform input dimensionality.

\subsection{Particle Dual Attention Transformer }

P-DAT takes the particle information within the jet as input and consists of three main components: (i) the feature extractor, (ii) the particle attention module, and (iii) the channel attention module. In this work, we introduce a quantized variant, P-DAT-Bit, where we apply different quantization strategies to various parts of the model.

Specifically, we employ QAT for the particle and channel attention modules while keeping other components, including the feature extractor, 1D CNN, and final multi-layer perceptron (MLP) classifier, in full precision. This choice is motivated by computational efficiency and robustness considerations. The attention modules contain the majority (\~63$\%$) of parameters and are the most computationally demanding, thus offering the greatest benefit from quantization. Furthermore, transformer-based attention mechanisms have been shown to exhibit notable robustness to quantization, making them an ideal starting point for a proof-of-concept study. Therefore, we implement QAT by replacing all linear layers with BitLinear layers in the two particle attention and two channel attention modules, affecting approximately 63\% of the total weight parameters. The attention modules incorporate physics-motivated bias terms in the scaled dot-product attention, which remain in full precision to preserve critical information. The final output undergoes a 1D CNN transformation and global average pooling before being fed into an MLP classifier. Further architectural details can be found in Ref.~\cite{He:2023cfc}. The new model is trained from scratch, with all hyperparameters identical to the non-quantized version.

\subsection{Performance comparison}

\begin{table}[t!]
    \centering
    \setlength{\tabcolsep}{10pt} 
    \begin{small}
	\begin{tabular}{lcccS[table-format=3.1(2)]}
     \toprule
    &  Accuracy & AUC & Rej$_{50\%}$  & {Rej$_{30\%}$} \\
    \midrule
    ParticleNet~\cite{Qu:2019gqs} & $0.840$ & $0.9116$ & $39.8(2)$ & 98.6(13)\\
    PCT~\cite{Mikuni:2021pou} & 0.841 & 0.9140 & $43.2(7)$ & 118.0(22) \\
    LorentzNet~\cite{Gong:2022lye} & 0.844 & 0.9156 & $42.4(4)$ & 110.2(13) \\
    ParT \cite{Qu:2022mxj} & 0.849 & 0.9203 & $47.9(5)$ & 129.5(9) \\
    \midrule
    P-DAT \cite{He:2023cfc} & 0.839 & 0.9092 & $39.2(6)$ & 95.1(13) \\
    \midrule
    P-DAT-Bit & 0.834 & 0.9040 & $35.0(3)$ & 83.3(12) \\
    \bottomrule
	\end{tabular}
    \end{small}
    \caption{Performance comparison for P-DAT, P-DAT-Bit, and some existing classification algorithms on the quark-gluon discrimination dataset. The uncertainties on rejection rates are calculated by taking the standard deviation of 5 independent training runs. The error bars for the accuracy and the AUC are not displayed as they are negligible.}
    \label{tab:results_qg}
\end{table}

As depicted in Tab.~\ref{tab:results_qg}, while P-DAT itself offers robust performance with an accuracy of 0.839 and an AUC of 0.9092, the adaptation of this model into P-DAT-Bit reveals both the strengths and the trade-offs of utilizing BitLinear layers within the attention blocks. The integration of BitLinear layers in P-DAT-Bit results in a slight decrease in accuracy and AUC compared to the non-quantized variant. Despite this modest drop, the performance metrics remain highly competitive, indicating that the reduced precision computation does not drastically compromise the model's discriminative capability. Specifically, the background rejection rates at 50\% signal efficiency (Rej$_{50\%}$) and 30\% signal efficiency (Rej$_{30\%}$) are 35.0 and 83.3, respectively. This variation between P-DAT and P-DAT-Bit highlights the balancing act between computational efficiency and performance accuracy. It emphasizes that while BitLinear layers streamline model operations,particularly advantageous in resource-constrained environments, there is a nuanced impact on the model’s ability to manage complex discriminative tasks.

\begin{figure}[b!]
    \centering
    \includegraphics[width=0.66\linewidth]{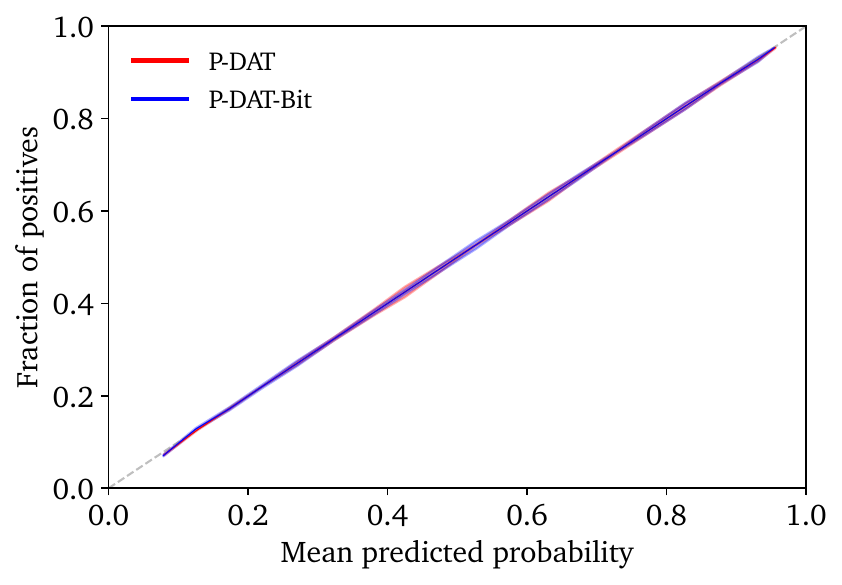}
    \caption{Calibration curves of P-DAT and P-DAT-Bit models for quark/gluon discrimination. The errorband has been obtained from evaluating 5 independent runs.}
    \label{fig:pdat_calib}
\end{figure}

Finally, in Fig.~\ref{fig:pdat_calib}, we present the calibration curves of P-DAT and P-DAT-Bit models for quark/gluon discrimination. The red line represents the calibration curve for the original P-DAT model, while the blue line represents the P-DAT-Bit model. Each model was evaluated over five runs to compute the average calibration curve and the associated standard deviation. Both models demonstrate good calibration performance, with their curves closely following the diagonal line, indicating that the predicted probabilities are well-calibrated with the actual positive class probabilities. Notably, the error bands for both models are very small, suggesting that the predictions of both models are consistent across multiple runs. Besides, the blue band of the P-DAT-Bit model is slightly broader compared to the red band of the P-DAT model, which can be attributed to its use of lower bit precision for weights and inputs in 60\% of its parameters. Despite this, the P-DAT-Bit model is well calibrated and closely following the ideal diagonal line, indicating that its predicted probabilities are generally accurate.

To quantify the computational cost of the P-DAT-Bit, we compute the number of FLOPs, IntOPs, and SignOPs of its two building blocks, the Particle Attention Block and the Channel Attention Block in Tab.~\ref{tab:comparison_pdat}. 

\begin{table}[t!]
\centering
{\small
\setlength{\tabcolsep}{10pt}  
\begin{tabular}{c|l S[table-format=7.0] S[table-format=7.0]@{\;}l}
\toprule
Block & Operations & {P-DAT} & \multicolumn{2}{c}{P-DAT-Bit} \\
\midrule
\multirow{4}{*}{\makecell[c]{Particle\\Attention\\Block}} 
&FLOPs & 7795200 & 2995590 &\\
&SignOPs & {-} & 2457600 &\\
&IntOPs & {-} & 2419200 &\\
&Total FLOPs & 7795200 & 3237510 &(41\%) \\
\midrule
\multirow{4}{*}{\makecell[c]{Channel\\Attention\\Block}} 
&FLOPs & 5896192 & 1896517 &\\
&SignOPs & {-} & 2048000 &\\
&IntOPs & {-} & 1209600 &\\
&Total FLOPs & 5896192 & 2017477 &(34\%)\\
\bottomrule
\end{tabular}
}
\caption{Computational cost for a single pass of the full-precision and \bitnet implementations of the Particle Attention and Channel Attention Blocks in the P-DAT architecture. Percentages are in relation to the number of FLOPs of P-DAT.
Assumptions: 1 FLOP = 10 IntOPs, SignOP = 0 FLOPs.}
\label{tab:comparison_pdat}
\end{table}

\clearpage
\section{Regression: SMEFT parameter estimation}
\label{sec:regression}

In this section, we evaluate the performance of \bitnet for a regression task and focus on the SMEFTNet~\cite{Chatterjee:2024pbp} architecture. We denote its quantized version as SMEFTNet-Bit, where either all or some linear layers are replaced with its BitLinear counterpart. To perform a meaningful comparison, we employ the same simulated $\PW\PZ$ event samples used in Ref.~\cite{Chatterjee:2024pbp}, focusing on a semi-leptonic decay chain
\begin{align}
    \Pp\Pp \to \PW(\to\Pq\bar\Pq)\,\PZ(\to\ell\bar\ell)\;.
\end{align}
The goal is to predict the decay plane angle $\phi^\text{jet}_\text{decay}$ of the parton-level quarks from the particle-level jet information, as this observable is sensitive to linear SMEFT–SM interference effects. This decay plane angle depends on the exact momenta of the up-type and down-type quarks, and interchanging these momenta at the parton level results in a difference of $\pi$. However, the particle-level decay products do not contain this information and are thus invariant under this permutation. Hence, in order to capture this ambiguity in the training of the network $f_\theta(x)$ with trainable parameters $\theta$, we use the modified loss function~\cite{Chatterjee:2024pbp}
\begin{align}
\mathcal{L}=\left\langle\sin^2 \left(f_\theta(x_i)-\phi^\text{jet}_{i,\text{decay}}\right)\right\rangle_{(x_j,\varphi_j)\sim\mathcal{D}_\text{train}}\;,
\end{align}
where the $\sin$ encodes the symmetry of shifting with $\pi$.
The input $x$ consists of the lab-frame momenta of each particle in a given event $j$ parametrized as
\begin{align}
    x=\{p_{\text{T},1},\varphi_1,\Delta R_1,\cdots, p_{\text{T},N_j},\varphi_{N_j},\Delta R_{N_j}\}\;,
\end{align}
where $N_j$ denotes the number of particles in this event.

\subsection{SMEFTNet architecture}

SMEFTNet is an IRC-safe and rotation-equivariant graph neural network. It is designed to provide an optimal observable for small deviations from the SM and enhance SMEFT sensitivity, specifically focusing on the linear SM-SMEFT interference within SMEFT. To preserve sensitivity to the linear term, it is crucial to consider the orientation of the decay planes of the \PW~or \PZ~boson, as this orientation helps resolve the helicity configuration of the amplitude that is altered in the SMEFT~\cite{Panico:2017frx}. To study the hadronic final states of \PW~or \PZ~boson, SMEFTNet is constructed to be equivariant to azimuthal rotations of the constituents of the boosted jet around the jet axis, maintaining SO(2) symmetry regardless of the chosen reference frame. It processes inputs as variable-length lists of particle constituents of a fat jet, originating from the hadronic decay of a boosted massive particle. For details on the architecture, we refer to Ref.~\cite{Chatterjee:2024pbp}. Although the decay-plane angle regression is a relatively lightweight task, investigating quantization in this controlled setting is still meaningful. In realistic SMEFT inference tasks, surrogate models have to be evaluated repeatedly when profiling over many nuisance parameters to estimate systematic uncertainties, or when integrating over latent observables, where lighter and faster models become highly advantageous, and even moderate speedups can lead to substantial reductions in end-to-end inference time. Thus, assessing whether low-precision variants of lightweight models retain accuracy is a necessary first step before applying such techniques in more computationally intensive inference workflows.

In our study, we consider three quantized variants of the SMEFTNet model and apply the three variants to predict the decay plane angle from the jet’s constituents and compare their performance with the results presented in Ref~\cite{Chatterjee:2024pbp}. In the first variant, all linear layers are replaced with BitLinear layers, which we refer to as SMEFTNet-Bit100. In the second variant, only the linear layers within the MLP block are quantized, corresponding to approximately 70\% of the total weight parameters; this configuration is labeled SMEFTNet-Bit70. In the third variant, only the linear layers in the Message Passing Neural Network block are quantized, accounting for about 30\% of the weights; this is denoted SMEFTNet-Bit30.

\subsection{Performance comparison}

\begin{table}[t!]
    \centering
    \setlength{\tabcolsep}{10pt} 
    \begin{small}
	\begin{tabular}{lccccc}
     \toprule
Model Comparison  & Wasserstein Distance & Separation Power \\
    \midrule
    SMEFTNet & 0.0021 & 0.0001 \\    
    SMEFTNet-Bit100 & 0.4546 & 1.4860 \\
    SMEFTNet-Bit70 & 0.2528 & 0.6138 \\
    SMEFTNet-Bit30 & 0.1040 & 0.1503 \\
    \bottomrule
	\end{tabular}
    \end{small}
    \caption{Comparison of the Wasserstein distance and separation power between the residual distributions of each of the three quantized SMEFTNet variants, an independently re-trained full-precision SMEFTNet, and that of the original SMEFTNet.}
    \label{tab:results_smeft}
\end{table}

\begin{figure}[b!]
    \includegraphics[width=0.49\linewidth]{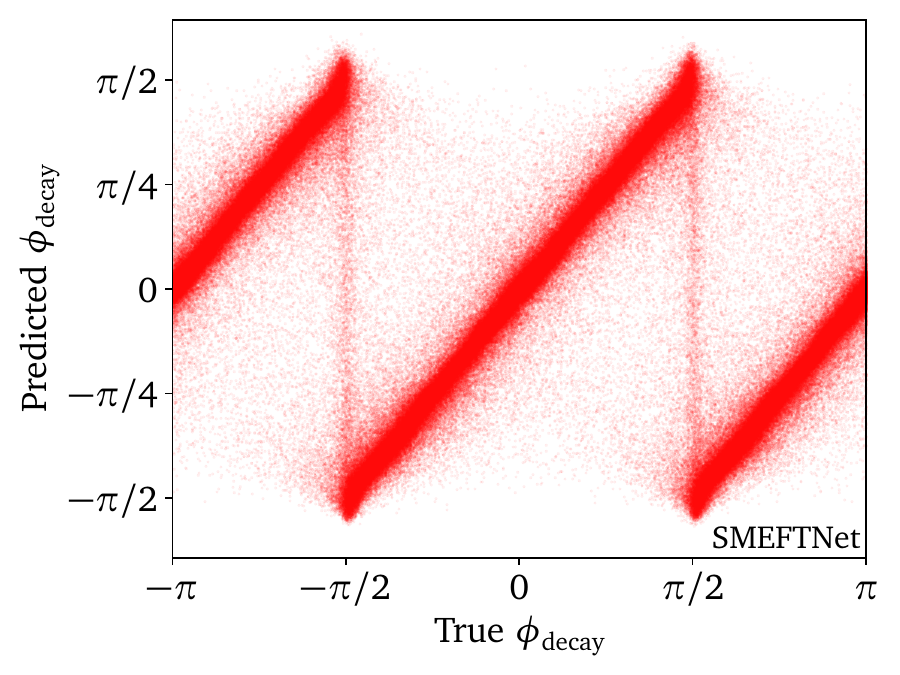}
    \includegraphics[width=0.49\linewidth]{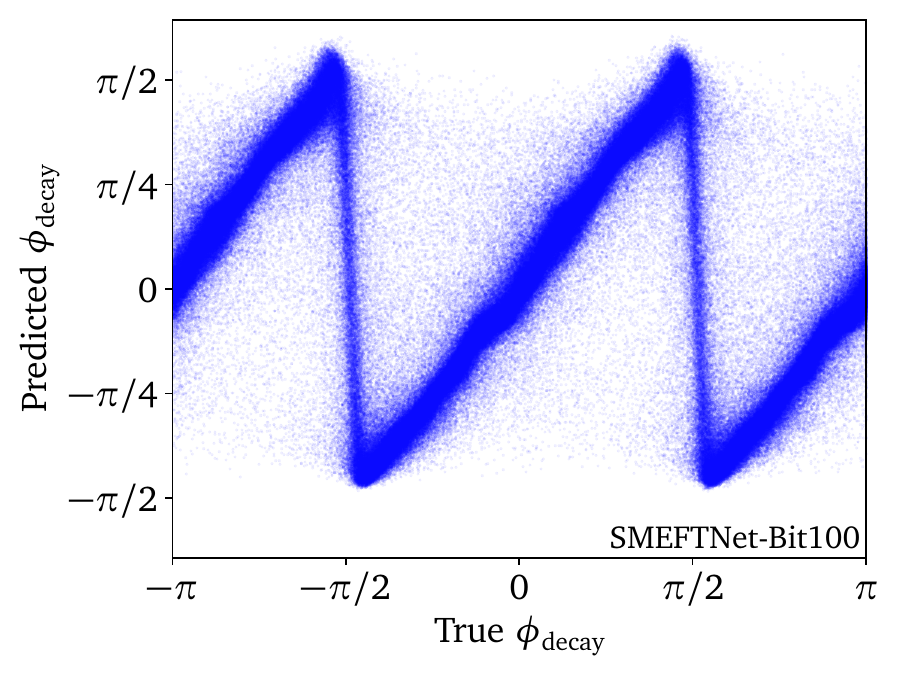}
  \caption{Two-dimensional scatter plots of the true and predicted angle $\phi_{\text{decay}}$ in the test dataset. Left: SMEFTNet with regular linear layers. Right: SMEFTNet-Bit100 with BitLinear layers.}
  \label{fig:scatter}
\end{figure}

To visualize the comparative analysis of SMEFTNet and SMEFTNet-Bit100, Fig.~\ref{fig:scatter} presents the scatter plots of the true and the predicted $\varphi_{\textrm{decay}}$ in the remaining 20\% $\PW\PZ$ dataset, respectively. The results reveal that while the SMEFTNet-Bit100 model marginally underperforms relative to the original SMEFTNet, the differences are minimal, showcasing the effectiveness of the low-bit model in capturing the essential structure of the data despite its limited weight and input precision. Notably, the scatter plot of SMEFTNet exhibits two faint vertical red bands at true $\phi_\text{decay}$ values of $\pm\pi/2$, which become more pronounced and broader in the SMEFTNet-Bit100 scatter plot. This phenomenon is attributed to both models' tendency to stabilize at local minima due to the inherent challenges posed by multi-objective optimization. In the critical region of true $\phi_{\text{decay}} \approx \pm \pi/2$, two physically indistinguishable parton-level configurations — arising from the permutation of up-type and down-type quarks — correspond to decay plane angles that differ by $\pi$. Since the particle-level information does not resolve this ambiguity, the regression model is trained with both values as plausible targets. As a result, it is forced to compromise between conflicting objectives, potentially leading to suboptimal predictions for either configuration individually. 
The broader bands in the SMEFTNet-Bit100 results are explained by its limited precision due to weight and input quantization. This restricted precision leads to significant fluctuations in the predicted $\phi_{\text{decay}}$ values, resulting in the observed wider bands in the visual representations.

\begin{figure}[bt!]
    \includegraphics[width=0.33\linewidth]{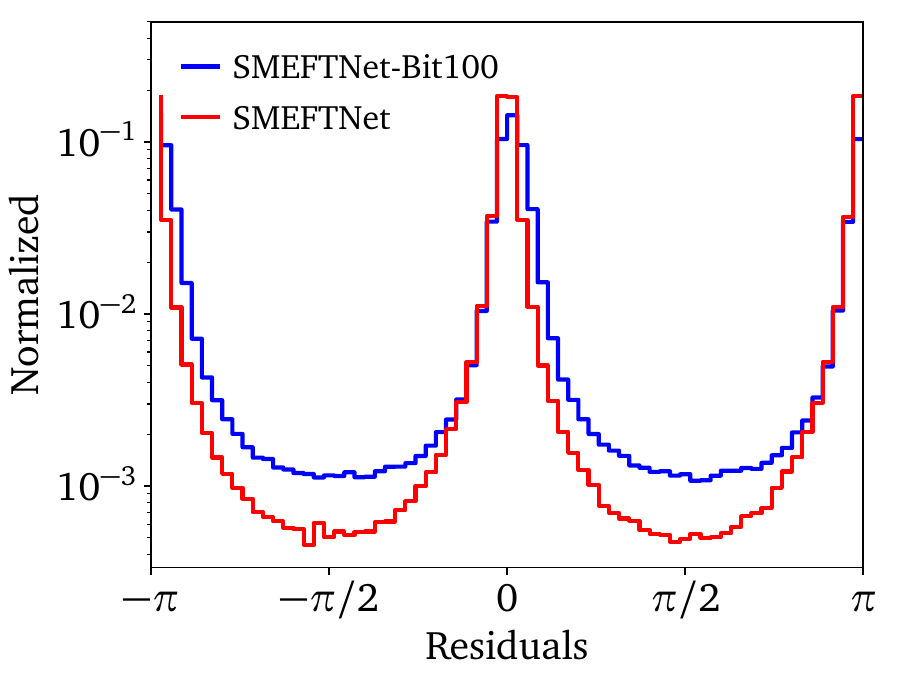}
    \includegraphics[width=0.33\linewidth]{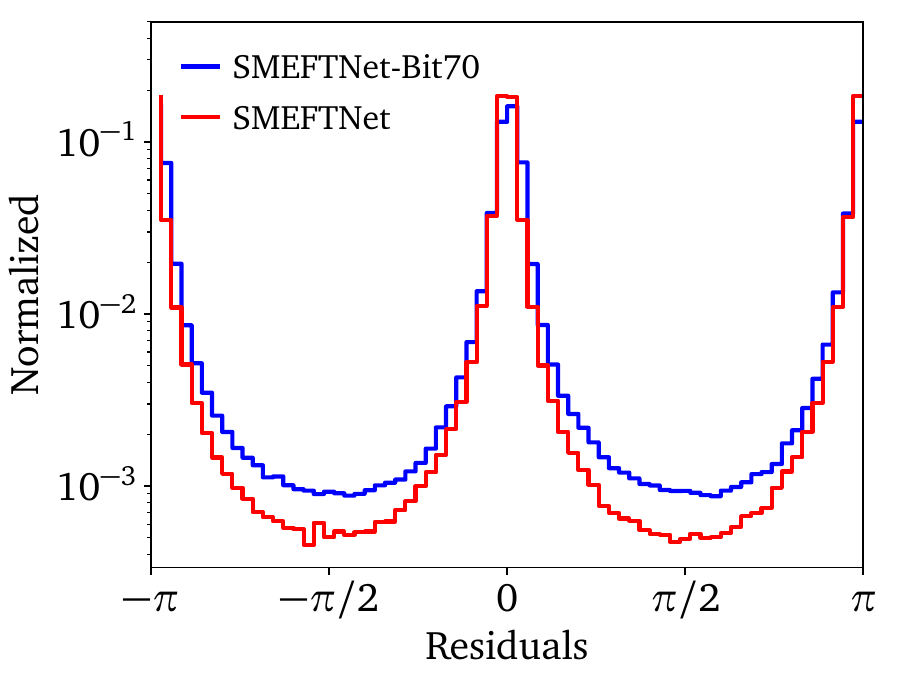}
    \includegraphics[width=0.33\linewidth]
    {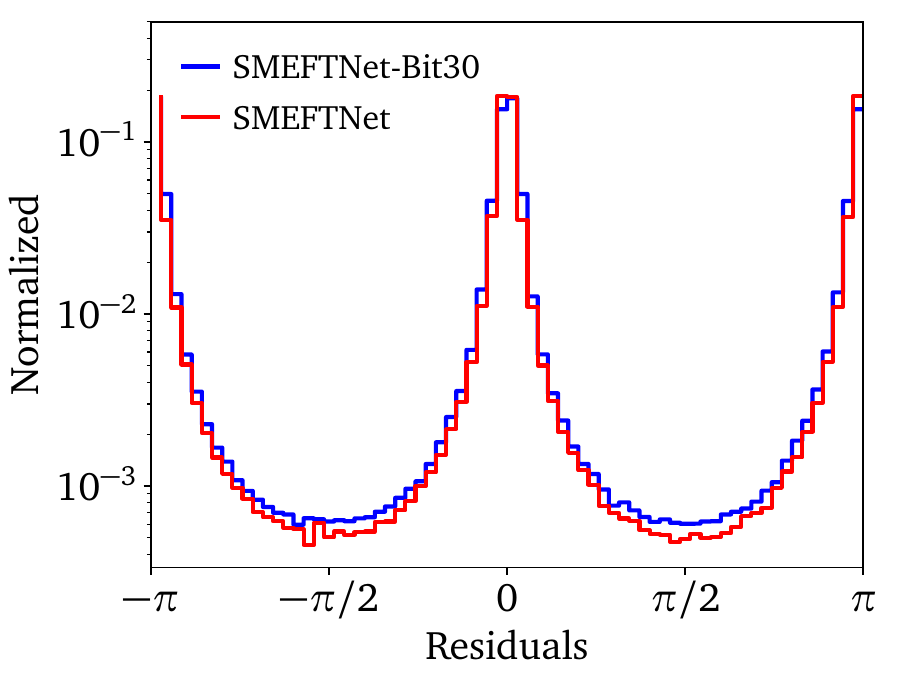}
    \caption{Histograms of residuals, defined as the differences between truths and predictions for $\phi_\text{decay}$ regression task. The histogram in blue represents SMEFTNet without BitLinear layers, whereas the histogram in red corresponds to SMEFTNet with BitLinear layers. From left to right: 100\%, 70\%, and 30\% of the weights are quantized.}
    \label{fig:comparison}
\end{figure}

Finally, Fig.~\ref{fig:comparison} presents three histograms of residuals, which represent the differences between truths and predictions for $\phi_\text{j, decay}$. In each histogram, the red bars correspond to the original SMEFTNet, while the blue bars represent different configurations of the SMEFTNet-Bit model. In the left histogram, the blue bars correspond to SMEFTNet-Bit100. In the middle histogram, the blue bars represent the SMEFTNet-Bit70. In the right histogram, the blue bars represent the SMEFTNet-Bit30. In all cases, the SMEFTNet model exhibits a sharp peak at zero, indicating a precise alignment of predictions with true values. When fully quantized (100\%), the SMEFTNet-Bit100 distribution broadens prominently, reflecting larger variations in prediction accuracy. Reducing the fraction of quantized layers to 70\% yields intermediate performance: the residuals of SMEFTNet-Bit70 remain somewhat more dispersed than SMEFTNet but are considerably narrower than the fully quantized version. At 30\% quantization, the residuals of SMEFTNet-Bit30 closely resemble those of the original SMEFTNet, showing only a modest increase in width. Consequently, it is evident that as more linear layers are replaced by \bitnet layers, the performance of the model deteriorates. Moreover, a notable feature in all plots is the periodic structure of the residuals, which arises because the model faces a periodic ambiguity in which angles separated by \(\pi\) map to the same predicted value, creating these peaks around $\pm \pi$ and 0 in the residual distribution. Furthermore, the fraction at \(\pm \frac{\pi}{2}\) is higher for SMEFTNet-Bit100, reflecting its broader vertical band in its 2D scatter plots. This broadening stems from the limited precision of BitLinear layers, which amplifies prediction fluctuations near critical values of $\pm \pi$. Notably, as the proportion of quantized layers increases, the fraction at \(\pm \frac{\pi}{2}\) also becomes higher, indicating that stronger quantization further destabilizes the predictions near critical values of $\pm \pi$. By contrast, SMEFTNet shows narrower bands, indicating more accurate predictions and a lower fraction at \(\pm \frac{\pi}{2}\). Lastly, across all three configurations, both models overlap around zero residuals, highlighting their overall reliability. However, the progressive widening of SMEFTNet-Bit’s distributions with increasing quantization underscores the trade-off between model compression and predictive precision. This demonstrates that partial, rather than complete, quantization can give a better balance for resource-limited applications. 

Table~\ref{tab:results_smeft} complements Fig.~\ref{fig:comparison} with two quantitative metrics. The Wasserstein distance captures the average angular shift between residual distributions, whereas the separation power reflects their overall shape divergence. The additional “SMEFTNet” entry corresponds to a second model trained with the same architecture and hyper-parameters as the reference model but initialized with a different random seed. Its near-zero Wasserstein distance (0.002) and separation power ($1\times10^{-4}$) serve as a self-consistency check, confirming that both metrics are sensitive only to distributional shifts and robust to statistical fluctuations. By contrast, the three quantized variants show progressively larger values as the proportion of BitLinear layers increases, quantitatively substantiating the broadening already visible in Fig.~\ref{fig:comparison}.

\begin{table}[t!]
\centering
{\small
\setlength{\tabcolsep}{12pt}  
\begin{tabular}{c|lS[table-format=4.0] S[table-format=4.0]@{\;}l}
\toprule
Block & Operations & {SMEFTNet} & \multicolumn{2}{c}{SMEFTNet-Bit100} \\
\midrule
\multirow{4}{*}{\makecell[c]{Message\\Passing\\Block}} 
&FLOPs & 1080 & 188 &\\
&SignOPs & {-} & 520 &\\
&IntOPs & {-} & 479 &\\
&Total FLOPs & 1080 & 288 &(27\%) \\
\midrule
\multirow{4}{*}{\makecell[c]{MLP\\Block}} 
&FLOPs & 3776 & 411 &\\
&SignOPs & {-} & 1824& \\
&IntOPs & {-} & 1759 &\\
&Total FLOPs & 3776 & 769 &(20\%)\\
\bottomrule
\end{tabular}
}
\caption{Computational cost for a single pass of the Full-Precision and \bitnet implementations of the Message Passing and MLP blocks in the SMEFTNet architecture. Percentages are in relation to the number of FLOPs of the SMEFTNet. Assumptions: 1 FLOP = 10 IntOPs, SignOP = 0 FLOPs.}
\label{tab:comparison_smeftnet}
\end{table}

To estimate the resource consumption of the SMEFTNet-Bit100, we compute the number of FLOPs, IntOPs, and SignOPs of its two building blocks, the Message Passing Block and the MLP Block in Tab.~\ref{tab:comparison_smeftnet}. Since the Message Passing Block is applied once per edge while the MLP Block is applied once per event, we report their per-pass costs separately. 

\clearpage
\section{Generative: Detector simulation}
\label{sec:gen_exp}

Next, we consider a generative task. These are very important in HEP, as the process of sampling random events from a given (complicated) conditional probability density is precisely what happens in numerical simulations of particle collisions (See~\cite{Butter:2022rso} for an overview on ML applications to the entire simulation chain.). One notable challenge is detector simulations, which involve modeling showers of energetic particles in the calorimeters, a task characterized by its high numerical complexity and dimensionality. In the past years, this field has seen many new ideas and approaches~\cite{Hashemi:2023rgo}, which led to the conception of the CaloChallenge~\cite{Krause:2024avx} to compare existing models on equal footing and spur even more development of state-of-the-art methods. 
The CaloChallenge was a data challenge in the HEP community, with four different datasets, increasing in their dimensionality. The goal of the challenge was to train generative networks on the datasets and to generate artificial samples as fast and precise as possible~\cite{calochallenge}. The dataset dimensionalities range from a few hundred in the easiest case to a few tens of thousands in the most complicated case. These are single particle showers, simulated with \geant~\cite{Agostinelli:2002hh,1610988,ALLISON2016186} and available at~\cite{CaloChallenge_ds1,CaloChallenge_ds1_v3,CaloChallenge_ds2,CaloChallenge_ds3}. For more details on the datasets we refer to~\cite{Krause:2024avx}.

\begin{figure}[bp!]
    \centering
    \includegraphics[page=3, width=0.95\linewidth]{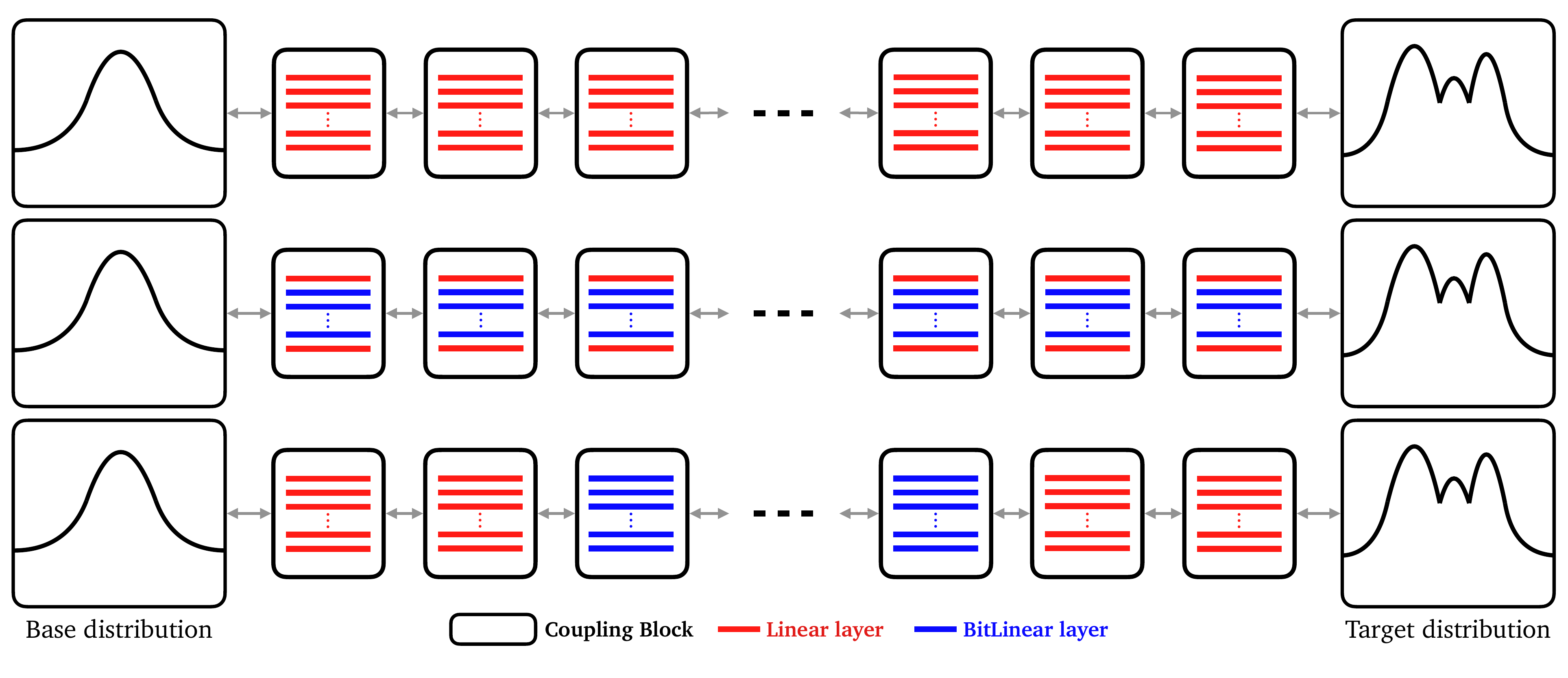}
    \caption{Illustration of different levels of quantization for the flow. {\color{red}Red} lines represent regular linear layers, while {\color{blue}blue} lines indicate BitLinear layers. The three setups, from top to bottom, are: \emph{regular}, \emph{NNCentral}, and \emph{BlockCentral}.}
    \label{fig:quantized_flow}
\end{figure}

When evaluating generative models, one can pick from various metrics~\cite{Kansal:2022spb,Das:2023ktd,Krause:2024avx}. Here, we use classifiers to evaluate the generative performance. The classifiers are trained on the task of distinguishing generated samples from the \geant reference. If a powerful, well-trained classifier cannot distinguish between the two samples, we conclude that the samples were drawn from the same underlying distribution, \ie the generative model learned the underlying distribution well~\cite{Krause:2021ilc}. Summarized in a single quantity, we use the area under the receiver operating characteristic (AUC). A lower AUC indicates a better generative model.

As an example for generative networks, we consider two well-performing submissions of the CaloChallenge~\cite{Krause:2024avx}: \caloinn~\cite{Ernst:2023qvn} based on a normalizing flow and \calodream~\cite{Favaro:2024rle} based on conditional flow matching. Normalizing flows were the first generative model that passed the ``classifier test''~\cite{Krause:2021ilc,Das:2023ktd} in calorimeter shower simulations and have seen various applications to this task~\cite{Buss:2024orz,Ernst:2023qvn,Buckley:2023rez,Pang:2023wfx,Schnake:2024mip,Diefenbacher:2023vsw,Krause:2021ilc,Krause:2021wez,Krause:2022jna,Krause:2024avx}. Conditional flow matching is the most recent generative model that was explored in this context and shows impressive performance in many applications~\cite{Favaro:2024rle,Cresswell:2024fst,Dreyer:2025zhp}.

\subsection{Normalizing Flow -- \caloinn}

\caloinn~\cite{Ernst:2023qvn} is a normalizing-flow-based generative model, learning a bijective transformation between a simple base distribution, usually a Gaussian, and a more complicated distribution, the target distribution. In detail, \caloinn is based on coupling layer-based normalizing flow~\cite{Dinh2014NICE,Dinh:2016pgf} (which are sometimes also called INN, invertible neural networks~\cite{ardizzone2019analyzinginverseproblemsinvertible}, even though they all are invertible). These types of flows are equally fast to evaluate in both directions (as density estimator and generative model), making training and generation more efficient than in an autoregressive setup~\cite{Krause:2021ilc,Krause:2021wez,Krause:2022jna}. The bijective transformation that is used in the coupling blocks is based on splines~\cite{mueller2019neural_importance_sampling,durkan2019neural_spline_flows,durkan2019cubic}. 

The calorimeter showers are normalized to unit energy in each calorimeter layer, and the corresponding layer energies are then appended to this array and encoded via ratios to the incident energy. This setup allows \caloinn to learn the distribution of calorimeter showers in a single step instead of a more time-consuming two-step procedure~\cite{Krause:2021ilc,Krause:2021wez,Krause:2022jna}. The downside of this approach is the scaling with the dimensionality of the dataset. The spline-based transformation requires a large number of output parameters for the individual NNs in the coupling layers, which in turn increases the number of trainable parameters of the NN. For dataset 2, \caloinn already has $270$M parameters, making an application to the six times bigger dataset 3 of the CaloChallenge impossible. 

All hyperparameters are as in the original publication~\cite{Ernst:2023qvn}, with the flow consisting of 12 (14) coupling blocks and the individual NNs having 256 hidden nodes, and 4 (3) hidden layers for dataset 1 (2). For dataset 1, \caloinn uses rational quadratic splines~\cite{durkan2019neural_spline_flows}, for dataset 2 it uses cubic splines~\cite{durkan2019cubic}\footnote{This choice was motivated in~\cite{Ernst:2023qvn} through an improved stability. Indeed, we also observe a few cases of NaNs being sampled: about 40 cases in the \emph{NNCentral} setup, 1 case in the \emph{BlockCentral}, and 4000 cases in the \emph{All} scenario.}. For quantization, we consider 5 different setups, or quantization strategies, which we can summarize as follows:
\begin{description}
  \item[Default:] The setup used in the original publication~\cite{Ernst:2023qvn}, without quantization. It serves as a baseline as well as a confirmation that we use the same hyperparameters as Ref.~\cite{Ernst:2023qvn}.

  \item[Exchange Permutation:] A setup where we change the permutations between the bijector blocks. Instead of using random permutations throughout, we exchange the sets that are transformed and that go into the NN after the first and before the last bijector. This strategy ensures that every dimension is transformed exactly once by both the first and last two bijectors. There is no quantization in this setup. It serves as a baseline for the \emph{BlockCentral} setup, as it shares the same permutation scheme.

  \item[NNCentral:] Only the central layers in each NN are quantized. Since in spline-based flows the final layer typically has the most parameters, this configuration has minimal effects.

  \item[BlockCentral:] Quantization is applied to all layers of NNs in the \emph{central} bijectors, \ie neither the first two nor the last two in the chain. Paired with the same permutation strategy as in \emph{Exchange Permutation}, this ensures each dimension is first transformed by a regular bijector, then by quantized ones, and finally again by a regular bijector.

  \item[All:] All linear layers in all bijectors are quantized.
\end{description}
Fig.~\ref{fig:quantized_flow} shows an illustration of the \emph{regular}, \emph{NNCentral} and \emph{BlockCentral}. Their fractions of quantized weights are given in Tab.~\ref{tab:results_caloinn}.
To evaluate the performance of the quantized \caloinn, we use the exactly same classifier architecture and training strategy as in \cite{Krause:2024avx}, making the results directly comparable. The main results are shown in Tab.~\ref{tab:results_caloinn}.

\begin{table}[htb]
    \centering
    \setlength{\tabcolsep}{5pt}
    \begin{small}
    \begin{tabular}{l c|c  S[table-format=2.1,table-space-text-post=\%] cc}
     \toprule
    Dataset & \multicolumn{2}{c}{Setup} & {Quantization} & Low-Level AUC & High-Level AUC \\
    \midrule
    \multirow{5}{*}{ds1--$\gamma$} 
    & \multirow{2}{*}{\rotatebox{90}{reg.}}& Default & {-} & 0.633(3) & 0.656(3) \\ 
    && Exchange Perm. & {-} & 0.640(4) & 0.651(3) \\
    \cmidrule{2-6}
    & \multirow{3}{*}{\rotatebox{90}{quant.}}& NNCentral & 8.4\% & 0.640(3) & 0.650(2) \\
    && BlockCentral & 66.6\% & 0.680(3) & 0.669(3) \\
    && All & 99.9\% & 0.759(2) & 0.828(2) \\
    \midrule
    \multirow{5}{*}{ds1--$\pi^+$} 
    & \multirow{2}{*}{\rotatebox{90}{reg.}}& Default & {-} & 0.793(3) & 0.742(3) \\  
    && Exchange Perm. & {-} & 0.784(2) & 0.736(3) \\
    \cmidrule{2-6}
    & \multirow{3}{*}{\rotatebox{90}{quant.}}& NNCentral & 5.9\% & 0.801(2) & 0.751(3) \\
    && BlockCentral & 66.6\% & 0.852(1) & 0.807(2) \\
    && All & 99.9\% & 0.882(2) & 0.907(2) \\
    \midrule
    \multirow{5}{*}{ds2} 
    & \multirow{2}{*}{\rotatebox{90}{reg.}}& Default & {-} & 0.738(4) & 0.859(2) \\  
    && Exchange Perm. & {-} & 0.728(6) & 0.857(3) \\
    \cmidrule{2-6}
    & \multirow{3}{*}{\rotatebox{90}{quant.}}& NNCentral & 0.3\% & 0.780(3) & 0.876(4) \\
    && BlockCentral & 71.4\% & 0.950(2) & 0.979(1) \\
    && All & 99.9\% & 0.993(1) & 0.998(0) \\
    \bottomrule
	\end{tabular}
    \end{small}
    \caption{Performance of default and \bitnet \caloinn using the classifier metric of the CaloChallenge~\cite{Krause:2024avx}. Uncertainties show the standard deviation over 10 random initializations and trainings of the classifier on the same \caloinn sample.}
    \label{tab:results_caloinn}
\end{table}

The AUCs we see in the regular setup are consistent with the ones reported in~\cite{Krause:2024avx}, where ds1 photon was scored 0.626(4) / 0.638(3), ds1 pion was scored 0.784(2) / 0.732(2), and ds2 was scored 0.743(2) / 0.865(3) for low / high-level observables, respectively. These AUCs differ from the ones presented in~\cite{Ernst:2023qvn} as the classifier architecture in the original publication was different than the one used here and in~\cite{Krause:2024avx}.   
The modified permutations usually improve the AUCs slightly, but mostly they agree with the regular setup within one standard deviation. 

\begin{table}[bh!]
\centering
\setlength{\tabcolsep}{10pt}
\begin{small}
\begin{tabular}{c|l S[table-format=8.0] S[table-format=8.0]@{\;}l S[table-format=8.0]@{\;}l}
\toprule
Dataset & Operations & {Default} & \multicolumn{2}{c}{NNCentral} & \multicolumn{2}{c}{All}\\
\midrule
\multirow{4}{*}{ds1--$\gamma$}
&FLOPs & 3121664 & 2861570 && 20982 & \\
&SignOPs & {-} & 131072 && 1560064 &\\
&IntOPs & {-} & 130560 && 1553902 &\\
&Total FLOPs & 3121664 & 2874626 & (92\%) & 176372 &(6\%) \\
\midrule
\multirow{4}{*}{ds1--$\pi^+$}
&FLOPs & 4411392 & 4151298 && 28373 &\\
&SignOPs & {-} & 131072 && 2204928&\\
&IntOPs & {-} & 130560 && 2196330 &\\
&Total FLOPs & 4411392 & 4164354 & (94\%) & 248006 &(6\%) \\
\midrule
\multirow{4}{*}{ds2}
&FLOPs & 38545408 & 38415361 && 220607 &\\
&SignOPs & {-} & 65536&& 19272704 &\\
&IntOPs & {-} & 65280 && 19200428&\\
&Total FLOPs & 38545408 &38421889 & (99.7\%) & 2140650 &(5.6\%)\\
\bottomrule
\end{tabular}
\end{small}
\caption{Computational cost for a single pass through one of the Coupling Blocks (CB) of the CaloINN. Percentages are in relation to the number of FLOPs of the regular CB. 
Assumptions: 1 FLOP = 10 IntOPs, SignOP = 0 FLOPs.}
\label{tab:comparison_caloINN}
\end{table}

We observe that quantizing the NN weights with the \bitnet degrades performance, with a clear correlation between the fraction of quantization and the AUC. Across all datasets, the \emph{NNCentral} setup gives AUCs almost as good as the \emph{Regular} setup, but in this case only a very low fraction of weights is actually quantized. The \emph{All} setup degrades the generative performance a lot, with the gap being larger in the high-dimensional dataset 2. The \emph{BlockCentral} setup results in an overall good performance. This is an interplay of not having quantized all of the parameters and the choice to only quantize the central coupling layers instead of the outer ones. This latter choice allows for the bulk of the bijection being carried out by a quantized flow, while the final details will be adjusted in the outer coupling blocks with a regular flow setup. 

To estimate the resource consumption of the quantized CaloINN, we compute the number of FLOPs, IntOPs, and SignOPs of the three different types of coupling layers that make up our quantization strategies in Tab~\ref{tab:comparison_caloINN}. While these number take into account the batchnormalizations in the NN, they do not include the FLOPs that are used in the spline evaluation. 

\subsection{Conditional Flow Matching -- \calodream}

\calodream~\cite{Favaro:2024rle} is a generative architecture combining Conditional Flow Matching (CFM) with transformer elements. 

In detail, \calodream consists of two networks, both trained with conditional flow matching: (i) an autoregressive transformer, called the \emph{energy network}, to learn the 45 layer energies, and (ii) a vision transformer, called the \emph{shape network}, learning the normalized showers. The self-attention mechanism in the transformer layers is highly beneficial for modeling the sparsity of calorimeter showers and the correlations across detector locations. Crucially, the introduction of \emph{patching}, which groups nearby voxels into coarser units, reduces the impact of the quadratic scaling and enables the model to scale efficiently to larger calorimeter geometries such as DS3. At the same time, conditional flow matching ensures efficient training of the underlying continuous normalizing flow model.

The main difference of \calodream to \caloinn when considering the quantization lies in the distribution of NNs in the generative process: instead of many ``smaller'' NNs in \caloinn, \calodream has one large NN for each of the two steps, so we expect the impact of quantization on performance to be smaller. 

The preprocessing of the calorimeter shower data is done as in the \caloinn setup. Showers are normalized per calorimeter layer, and the corresponding layer energies are encoded as ratios to constrain them to the range $\left[0,1\right]$. The energy network now learns the latter independently from the normalized showers, which are learned by the shape network. 

Since the two networks factorize, we can discuss quantization strategies separately and combine them ad libitum in generation. 

\subsubsection*{Quantization of the energy network}

In the energy network, an autoregressive transformer defines the embeddings of the energy values that are then passed to a MLP performing the CFM. We keep the original hyperparameter choices for these networks, so the former is based on 4 attention heads with an embedding dimension of 64, see~\cite{Favaro:2024rle}. The feed-forward CFM network consists of 8 hidden layers with 256 nodes each. It takes the concatenated 64-dimensional time embedding, the 64-dimensional autoregressive embedding of previous calorimeter layer energies as well as the 1-dimensional current calorimeter layer energy as input and predicts the velocity field of the current calorimeter layer as output. We consider two quantization setups for the energy network:
\begin{description}
  \item[Regular:] The non-quantized version, as used in the original publication~\cite{Favaro:2024rle}.

  \item[Quantized:] Quantizes the central 6 of the 8 hidden layers of the CFM network, leaving the transformer-based embedding networks untouched. This results in 66.09\% of the trainable parameters of the energy network being quantized. However, since the energy network is much smaller than the shape network, this translates to only 5.54\% of the total \calodream model being quantized.
\end{description}

\subsubsection*{Quantization of the shape network}

In the shape network, all 6480 voxels of the calorimeter are split into 135 patches of 48 voxels each, and all operations are performed on these patches in parallel. Diffusion time $t$, ratios of layer energies, incident energy, and voxel patches are transformed using three different embedding networks: (i) for time, (ii) for conditionals, and (iii) for position. These embeddings are passed to a chain of 6 Vision Transformer (ViT) blocks, each consisting of a self-attention layer, a projection, and a subsequent MLP step. The output is then passed through a final MLP layer and reassembled from patches into complete calorimeter showers.

Given the modular nature of the shape network, several quantization strategies are possible. Here, we report three:
\begin{description}
  \item[Regular:] The non-quantized version, identical to the setup in the original publication~\cite{Favaro:2024rle}.

  \item[No embedding:] Quantizes the core elements of the ViT blocks, \ie QKV matrices, projections, and MLPs, while keeping the embedding networks and the final MLP layer unquantized. This configuration results in 63.8\% of the shape network being quantized.

  \item[Full:] Extends the \emph{no embedding} setup by also quantizing the linear layers in the position, time, and conditional embedding networks. This increases the fraction of quantized parameters in the shape network to 66.22\%.
\end{description}

\subsubsection*{Performance comparison}

Also for the evaluation of \calodream, we use the same classifier architecture and training strategy as in the CaloChallenge~\cite{Krause:2024avx}, making the results directly comparable. The main results are shown in Tab.~\ref{tab:results_calodream}.

\begin{table}[htb]
    \centering
    \setlength{\tabcolsep}{5pt} 
    \begin{small}
	\begin{tabular}{ccS[table-format=2.1,table-space-text-post=\%]cc}
     \toprule
  Energy net & Shape net & {Quantization} & low-level AUC& high-level AUC \\
    \midrule
    regular & \multirow{2}{*}{regular} & {-} & 0.531(3) & 0.523(3) \\ 
    quantized && 5.5\% & 0.532(3) & 0.525(3) \\
    \midrule
    regular & \multirow{2}{*}{no embedding} & 58.5\% & 0.611(2) & 0.543(3) \\
    quantized && 63.9\% & 0.610(5) & 0.545(2) \\
    \midrule
    regular & \multirow{2}{*}{full} & 60.7\% & 0.735(4) & 0.942(2) \\
    quantized && 66.2\% & 0.738(4) & 0.944(3) \\
    \bottomrule
	\end{tabular}
    \end{small}
    \caption{Performance of regular and \bitnet \calodream using the classifier metric of the CaloChallenge~\cite{Krause:2024avx}. Uncertainties show the standard deviation over 10 random initializations and trainings of the classifier on the same \calodream sample. }
    \label{tab:results_calodream}
\end{table}

Our retraining of \calodream in the regular setup reproduces the scores of the CaloChallenge~\cite{Krause:2024avx} with 0.531(3) / 0.521(2) for low/high-level AUC. 

The first thing we observe in Tab.~\ref{tab:results_calodream} is that swapping the regular for the quantized energy network has no effect on the resulting AUC scores, so 66\% of trainable parameters in six out of eight hidden layers inside the CFM can safely be quantized without loss of sample quality.  

In the shape network, however, the performance strongly depends on which parts are quantized. A substantial part of the ViT blocks can be quantized with almost no loss in shower quality, but as soon as the embedding layers are quantized, the performance drops a lot. Note that the \emph{no embedding} \calodream, which is quantized to about 60\% still has the best high-level, and second best low-level AUCs of the CaloChallenge submissions~\cite{Krause:2024avx}. 

\begin{table}[t!]
\centering
{\small
\setlength{\tabcolsep}{10pt}
\begin{tabular}{c|l S[table-format=7.0] S[table-format=7.0]@{\;}l}
\toprule
Block & Operations & {Regular} & \multicolumn{2}{c}{Quantized}\\
\midrule
\multirow{4}{*}{\makecell[c]{Energy\\Network}}
&FLOPs & 3297292 & 163858 &\\
&SignOPs & {-} & 1572864 &\\
&IntOPs & {-} & 1569792 &\\
&Total FLOPs & 3297292 & 320837 &(10\%) \\
\bottomrule
\end{tabular}
}
\caption{Computational cost for a single pass through the energy network. Percentages are in relation to the number of FLOPs of the regular network. 
Assumptions: 1 FLOP = 10 IntOPs, SignOP = 0 FLOPs.}
\label{tab:comparison_caloDREAM_energy}
\end{table}

We also observe the spread between high- and low-level AUC to be larger. Especially in the \emph{fully} quantized case, the high-level features were already very sensitive. The low-level AUC score was not as bad, but that is likely a remnant of the classifier architecture choice, see the discussion in~\cite{Krause:2024avx}.

In tables \ref{tab:comparison_caloDREAM_energy} and \ref{tab:comparison_caloDREAM_shape} we give an estimate of the computational resources required for different components of \calodream. The numbers are based on a single run through the networks, so they only show a fraction of the FLOPs required to generate a full shower. In addition, we also do not include the transformer decoder and encoder networks of the energy network, as they are not quantized in our studies. 

\begin{table}[h!]
\centering
{\small
\setlength{\tabcolsep}{8pt}
\begin{tabular}{c|l S[table-format=8.0] S[table-format=8.0]@{\;}l S[table-format=8.0]@{\;}l}
\toprule
Block & Operations & {Regular} & \multicolumn{2}{c}{No embedding} & \multicolumn{2}{c}{Full}\\
\midrule
\multirow{4}{*}{\makecell[c]{Embedding\\Layers}}
&FLOPs & 1257600& 1257600 && 8515&\\
&SignOPs & {-} & {-} && 628800 &\\
&IntOPs & {-} & {-} && 626400 &\\
&Total FLOPs & 1257600 & 1257600 & (100\%)& 71155 &(6\%) \\
\midrule
\multirow{4}{*}{\makecell[c]{Diffusion\\Transformer\\Blocks}}
&FLOPs & 50771546 & 17691890 && 17691890 &\\
&SignOPs & {-} & 16588800 && 16588800 &\\
&IntOPs & {-} & 16562880 && 16562880 &\\
&Total FLOPs & 50771546 & 19348178 & (38\%) & 19348178 & (38\%) \\\bottomrule
\end{tabular}
}
\caption{Computational cost for a single pass through the shape network. Percentages are in relation to the number of FLOPs of the regular network. 
Assumptions: 1 FLOP = 10 IntOPs, SignOP = 0 FLOPs.}
\label{tab:comparison_caloDREAM_shape}
\end{table}

\clearpage
\section{Conclusions and Outlook}
\label{sec:outlook}

In this paper, we investigated the applicability of the \bitnet architecture to various tasks in high-energy physics. We demonstrated that QAT can offer a promising path forward for large-scale neural network applications in HEP. By applying these techniques to classification, regression, and generative tasks, we observed that performance remains competitive, especially for classification tasks such as quark–gluon tagging. However, the impact of quantization on regression and generative modeling is more nuanced and requires careful consideration of network size and architecture. The results highlight that:
\begin{itemize}
    \item \textbf{Larger networks can be quantized more easily.} Our experiments with different generative architectures used in detector simulation, specifically the normalizing flow-based \caloinn and the flow matching-based \calodream, indicate that larger models often exhibit smoother quantization behavior. For instance, with around 66\% of its parameters quantized, \caloinn experiences a noticeable drop in sample quality, whereas \calodream, which also quantizes roughly 63.8\% of its shape network, exhibits only minor performance degradation. These observations underscore how larger networks provide greater representational capacity, making them more resilient to the information loss introduced by low-bit representations. Consequently, these results highlight the potential of quantized, large-scale generative models in tackling the complex, high-dimensional tasks typical of modern particle physics experiments.
    
    \item \textbf{Performance depends on the layer chosen for quantization.} In our generative studies for detector simulation, \caloinn benefits from selectively quantizing all layers within the central bijectors (BlockCentral), thereby maintaining decent sample quality even at about 66\% quantized weights, while fully quantizing all layers (99.9\%) severely degrades performance. Likewise, in \calodream, quantizing the elements within the shape network’s ViT blocks (QKV matrices, projections, and MLPs) results in minimal performance loss; however, once the embedding layers are quantized, performance drops considerably. These findings underscore how carefully selecting which sections of the network to quantize -- central layers versus outer layers, embedding networks versus projection layers -- can preserve model fidelity and better adapt to each architecture’s demands, especially for high-dimensional tasks in particle physics. Further studies in automatic, heterogeneous quantization~\cite{Coelho:2020zfu,Sun:2024soe} are needed to fully extract the best performance of the networks. 
    
    \item \textbf{Fully quantized models show the largest performance degradation.} Our SMEFTNet experiments demonstrate that the fully quantized variant performs worst among all tested configurations. Specifically, SMEFTNet-Bit100 exhibits a larger degradation in accuracy than the configurations where only a subset of layers is quantized (SMEFTNet-Bit30 or SMEFTNet-Bit70). These results highlight the trade-off between model compression and accuracy, emphasizing the need for selective quantization to balance efficiency and precision.
    
    \item \textbf{Certain layers are robust to quantization.} Self-attention layers in transformer-based models, such as P-DAT or the shape network ViT in \calodream, exhibit surprisingly minimal performance degradation thanks to their ability to encode attention patterns effectively with discrete weights. This finding points to the potential of applying low-bit quantization in attention-driven architectures, enabling significant memory and computational savings.

    \item \textbf{Low-bit quantization aligns with future hardware and energy constraints.} As the HL-LHC generates increasingly large datasets which require more sophisticated analyses and simulations, achieving high performance while keeping the energy demands reasonable will be crucial. The low-bit QAT explored here aligns well with emerging hardware trends, where dedicated architectures for quantized operations are expected to play a central role.
\end{itemize}
The relevance of such quantization techniques is expected to grow as networks become larger and more prevalent throughout HEP workflows. Future research directions include exploring alternative weight quantization configurations, integrating these methods seamlessly into existing large-scale ML models, and adapting \bitnet for high-energy physics applications through fully quantized implementations with low-precision operations, enabling a quantitative evaluation of energy, memory, and latency reductions. Moreover, extending the scope to tasks requiring real-time or near-real-time performance will further validate the robustness of quantized models under realistic experimental conditions. A typical example is the LHC trigger system, which demands extremely fast classification on resource-limited hardware such as FPGAs. While quantization techniques have been successfully applied in this context for many years, our study highlights the potential of QAT to push these approaches further, beyond the trigger. By incorporating QAT, it may become feasible to implement larger and more expressive models on FPGAs or other hardware, potentially leading to more energy-efficient, scalable, yet accurate ML applications in HEP.

\section*{Acknowledgements}
We would like to thank Luigi Favaro, Ayodele Ore, and Sofia Palacios Schweitzer for assistance in setting up \caloinn and \calodream as well as Suman Chatterjee and Robert Schöfbeck for help with SMEFTNet. We would also like to thank Thea \AA{}rrestad for fruitful discussions on weight quantization. RW is supported 
through funding from the European Union NextGeneration EU program -- NRP Mission 4 Component 2 Investment 1.1 -- MUR PRIN 2022 – CUP G53D23001100006.  The computational results presented were obtained using the CLIP cluster (https://clip.science).

\clearpage
\bibliography{HEPML, other}

\end{document}